\documentclass[letterpaper, 10 pt, conference, right=0.6in, top= 0.701in]{IEEEtran}
\IEEEoverridecommandlockouts
\usepackage{cite}
\usepackage{amsmath,amssymb,amsfonts}
\usepackage{amsthm}
\usepackage{algorithmic}
\usepackage{tablefootnote}
\usepackage{graphicx}
\usepackage{textcomp}
\usepackage{tikz}
\usepackage{float}
\usepackage{xcolor}
\usepackage{multirow}
\usepackage{subfigure}
\usepackage{adjustbox}
\usepackage{stfloats}
\usepackage{balance}
\def\BibTeX{{\rm B\kern-.05em{\sc i\kern-.025em b}\kern-.08em T\kern-.1667em\lower.7ex\hbox{E}\kern-.125emX}}
\begin{document}

\title{{mmWave Communications for Indoor Dense Spaces: Ray-Tracing Based Channel Characterization and Performance Comparison}  }

\author{

 \IEEEauthorblockN{ Ozan Alp Topal\IEEEauthorrefmark{1}, Mustafa Ozger\IEEEauthorrefmark{1}, Dominic Schupke\IEEEauthorrefmark{2}, Emil Björnson\IEEEauthorrefmark{1}, and Cicek Cavdar\IEEEauthorrefmark{1}}

\IEEEauthorblockA{  \IEEEauthorrefmark{1} {School of Electrical Engineering and Computer Science,} {KTH Royal Institute of Technology}, Stockholm, Sweden  }

\IEEEauthorblockA{  \IEEEauthorrefmark{2} Airbus, Central Research and Technology,  Munich, Germany,}

\IEEEauthorblockA{E-mail: \IEEEauthorrefmark{1}\{oatopal, ozger, emilbjo, cavdar\}@kth.se), \IEEEauthorrefmark{2}dominic.schupke@airbus.com}
}
\maketitle

\begin{abstract}
In this paper, the indoor dense space (IDS) channel at 28 GHz is characterized through extensive Ray-Tracing (RT) simulations. We consider IDS as a specific type of  indoor environment with confined geometry and packed with humans, such as aircraft cabins and train wagons. Based on RT simulations, we characterize path loss, shadow fading, root-mean-square delay spread, Rician K-factor, azimuth/elevation angular spread of arrival/departure considering  different RT simulation scenarios of the fuselage geometry, material, and human presence. While the large-scale fading parameters are similar to the state-of-the-art channel models, the small-scale fading parameters demonstrate richer multipath scattering in IDS, resulting in poorer bit error rate performance in comparison to the 3GPP indoor channel model.
\end{abstract}


\section{Introduction}
The mobile network data traffic reached $72$ exabytes/month, while the 5G subscriptions with a 5G-capable device grew to $380$ million as of $2021$ \cite{ericsson}. A large proportion of the network traffic is generated by  emerging indoor applications, and within the new applications such as ultra high definition video streaming, and wireless cognition, the total demand of mobile network data is expected to grow  even further in 6G. To support the high demand, the millimeter wave (mmWave) bands (24 GHz-78 GHz) are promising in indoor applications due to the large amount of available bandwidth and relatively short distances.

Compared to the conventional sub-6-GHz bands, wireless signal propagation in the mmWave band is shown to be more sparse with less reflections and diffractions \cite{Rappaport_mmWave}. This effect makes the mmWave communication more sensitive to variations in the propagation environment and blocking of individual propagation paths. Therefore, the signal propagation has been modeled under different environment geometries, different objects inside the environment and their varying electromagnetic (EM) characteristics. In literature, mmWave communication channel has been extensively characterized for varying indoor applications such as residence, office, shopping mall, and factory \cite{Rappaport_mmWave, factory, cabin_60,train_28}.  Standard documents such as IEEE 802.11 ad/ay and 3GPP TR 38.901 classify and characterize different environments as site specific indoor models \cite{ieee_standard}, \cite{3gpp}.

In addition to the environments mentioned above, mmWave communication is also suitable for more densely populated environments, such as high-speed trains, airplanes, and (more recently) hyperloops. There are two main features of these environments that necessities to model them under a unified umbrella term. The first feature is that the considered space is smaller than other site-specific indoor environments, such as indoor offices, malls, or factories, and consequently the distances of radio nodes  are shorter. Second, these environments are filled up by many blocking objects and humans, which are generally static during several signaling intervals. In this paper, we term this specific type of environments as \textit{indoor dense spaces (IDS)}.

IDS come in different shapes, and flavours considering the objects composed of different materials. The geometry of the environment, and the geometry of the objects within is one of the varying properties among IDS that may strongly influence the signal propagation. While aircraft fuselage has a distinctive geometry similar to a half cylindrical surface, the train cabin is modeled with a rectangular shaped fuselage in \cite{train_60}. Even within the same IDS environment, the geometry of indoor objects such as overhead compartments and the seat positions might vary. Another important factor is the materials used for the objects.  Considering that the largest share of reflections and diffractions is caused by the body surrounding the IDS, its EM properties such as reflectively would alter the fading characteristics.  Finally, the most distinctive factor is the dense human presence in IDS. While the majority of the previous channel modeling works consider empty wagons or aircraft cabins in their measurements, passengers are also integral part of the environment, which may influence the channel characteristics. 

The literature on the mmWave channel characterization is based on channel sounding measurements in a specific environment that focus on the parameters of the fixed environment. For example, in \cite{train_60}, the authors provide statistical channel parameters of intra-wagon (IW) channels in high-speed trains considering the 60 and 300 GHz frequency bands. The only channel characterization work  considering the 28 GHz band for IW environments is \cite{train_28}, where the authors focus on LOS channel sounding measurements, and path loss and shadow fading characterization. For the aircraft channel modeling, \cite{cabin_60} conduct Ray-Tracing (RT) simulations, and provide received signal strength and coverage analysis. In \cite{cabin_measure_60}, the authors provide channel sounding measurements at 60 GHz. By providing lower path loss and functioning well in multipath environments, the 28 GHz mmWave band has propagation advantages in comparison to the 60 GHz mmWave bands \cite{Rappaport_mmWave}. However, wireless channel characterization for an in-cabin airplane environment at 28 GHz has not been considered in the literature. In IW environment, path loss and shadowing are modeled based on measurement campaign \cite{train_28}, but small-scale fading characteristics are not provided. Besides, dense human presence is also neglected in previous RT modeling and measurement campaigns. Instead of focusing on a single specific train or aircraft, in this work, we treat the considered environments as sub-categories of IDS environment, and model the propagation channel at 28 GHz under different geometry, material and denseness cases to illustrate the effects of these factors on the signal propagation. We conduct extensive RT simulations by taking into account material characteristics from the literature. The RT simulations provide realistic channel characteristics, where we can change the passenger existence and material characteristics while keeping everything else fixed. This provides the flexibility to analyze the effect of each variation unlike the measurement campaign works, where we cannot create different environments by keeping other properties fixed. The main contributions of this paper are as follows: 
\begin{itemize}
    \item We  propose IDS as a unified site-specific environment for trains and aircrafts, and model the signal propagation under different  scenarios based on considered materials, geometry and passenger existence.  According to the simulations, we analyze and discuss which channel parameters are unique for the IDS.
    \item This work is the first small-scale channel modeling effort considering the IW environments, and first full-scale channel modeling effort  at 28 GHz for aircraft cabins and hyperloops in the open literature.
    \item We provide path loss, root-mean-square ($\mathrm{RMS}$) delay spread, azimuth angle-of-spread departure ($\mathrm{ASD}$), azimuth angle of spread arrival ($\mathrm{ASA}$), elevation angle of spread departure ($\mathrm{ESD}$), elevation angle of spread arrival ($\mathrm{ESA}$), Rician K-factor ($\mathrm{KF}$) under different scenarios. We compare the results with a measurement campaign work to confirm the applicability of the RT model in the real time systems.
    \item We compare the bit error rate (BER) obtained with the proposed channel model with the 3GPP indoor office (3GPP IO) channel model.

\end{itemize}

\section{RT Modeling of Indoor Dense Spaces} 
In this section, we first detail the RT simulation environment and parameters for the considered simulation scenarios. We then provide the parameters regarding to the considered material characteristics and the link budget. 

\subsection{Simulation Scenarios}
\label{sec:sim_scenarios}
Precise modeling of the environment is critical for the mmWave channel since the signal propagation is sensitive to geometrical factors and materials. We use a commercial RT tool, Wireless Insite\footnote{Wireless InSite, available at: http://www.remcom.com/wireless-insite}, which provides reasonable channel modeling in complex environments by using the shooting-and-bouncing ray (SBR) model. To test the contribution of different factors in channel modeling, we will compare a baseline scenario with three different variations:
\begin{enumerate}
    \item Baseline (BL): We assume a perfect electric conductor (PEC) metal cylindrical  fuselage with all seats being occupied by passengers. This scenario corresponds to the hyperloop cabin. 
    \item {Composite Variation (C-V):} Here, we consider a carbon-glass composite cylindrical fuselage with all seats being occupied by passengers. A future aircraft cabin can be modeled by this scenario \cite{bachmann2017environmental}. 
    \item {Rectangular Variation (Rec-V):} In this scenario, we assume a rectangular PEC metal wagon cabin with all seats occupied by passengers. Train wagons are generally modeled by this type of scenario \cite{train_60, metro_28}. 
    
    \item {Empty Cabin Variation (Em-V):} We consider a PEC metal cylindrical fuselage without any passengers. This simulation scenario is important to demonstrate the impact of the dense user population in IDS.
\end{enumerate}

Each scenario corresponds to a variation of single parameter, while other parameters are fixed.  The remaining parameters for the TX-RX nodes are given in Table \ref{tab:link_budget}. The parameters in the last three rows of the table is the RT simulation configurations. 
\begin{table}[tb]
	\centering 
	\caption{Simulation Parameters for the RT.}
	\begin{tabular}{|l|c|}
		\cline{1-2}
		TX power           & 20 dBm    \\ \cline{1-2}
		Antenna gain (TX-RX)                    & 0 dBi    \\ \cline{1-2}
		Receiver noise figure        & 10 dB      \\ \cline{1-2}
		Bandwidth          & 1 GHz   \\ \cline{1-2}
		Carrier frequency          & 28 GHz   \\ \cline{1-2}
		Antenna polarization  & V-V (TX-RX)      \\ \cline{1-2}
		Antenna type                 & Isotropic    \\ \cline{1-2}
		Transmission line loss       & 0 dBm   \\ \cline{1-2} 
		Propagation mechanisms  & 3 reflection, 1 diffraction    \\ \cline{1-2}
		Diffuse scattering  & Enabled 1 reflection, 1 diffraction    \\ \cline{1-2}
		Scattering model  & Lambertian, coefficient= 0.4    \\ \cline{1-2}
	\end{tabular}
	\label{tab:link_budget}
\end{table}

\subsection{IDS Geometry}
\begin{figure}[tb]
	\begin{center}	
		\subfigure[]{
			\label{fig:north}
			\includegraphics[width=0.85\linewidth]{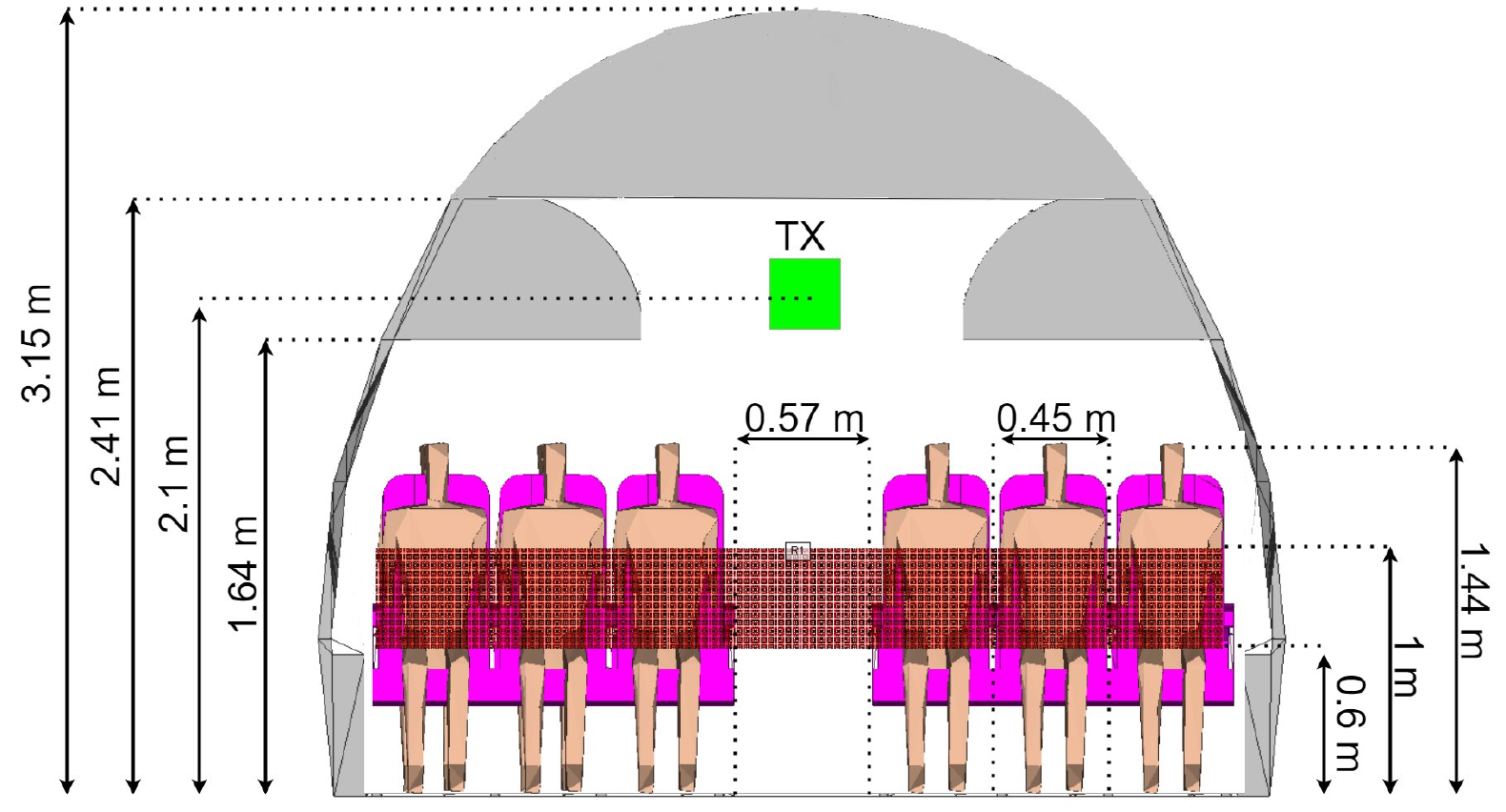} }\\
			\subfigure[]{
			\label{fig:aircraft_shell}
			\includegraphics[width=0.45\linewidth]{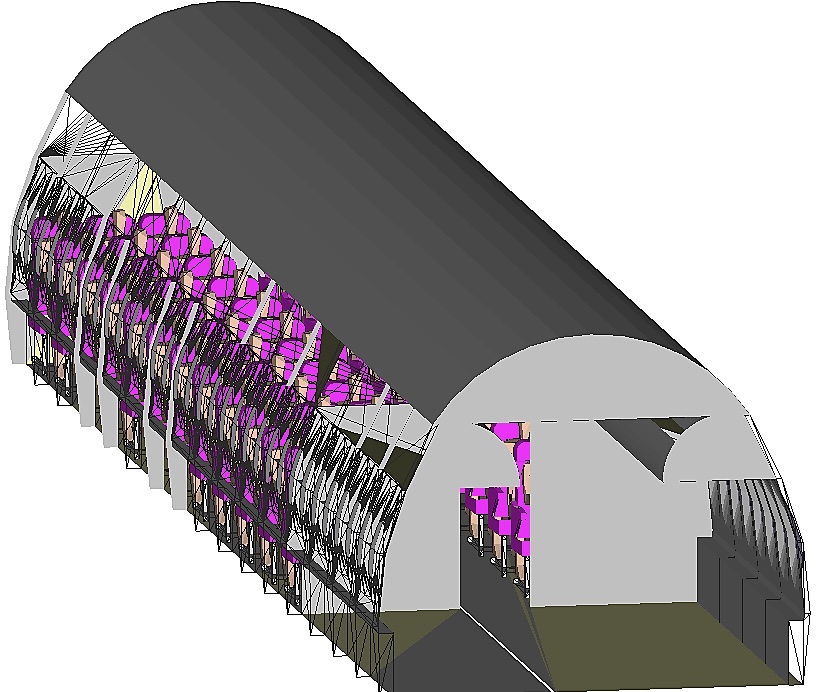} }	
		\subfigure[]{
			\label{fig:east}
			\includegraphics[width=0.45\linewidth]{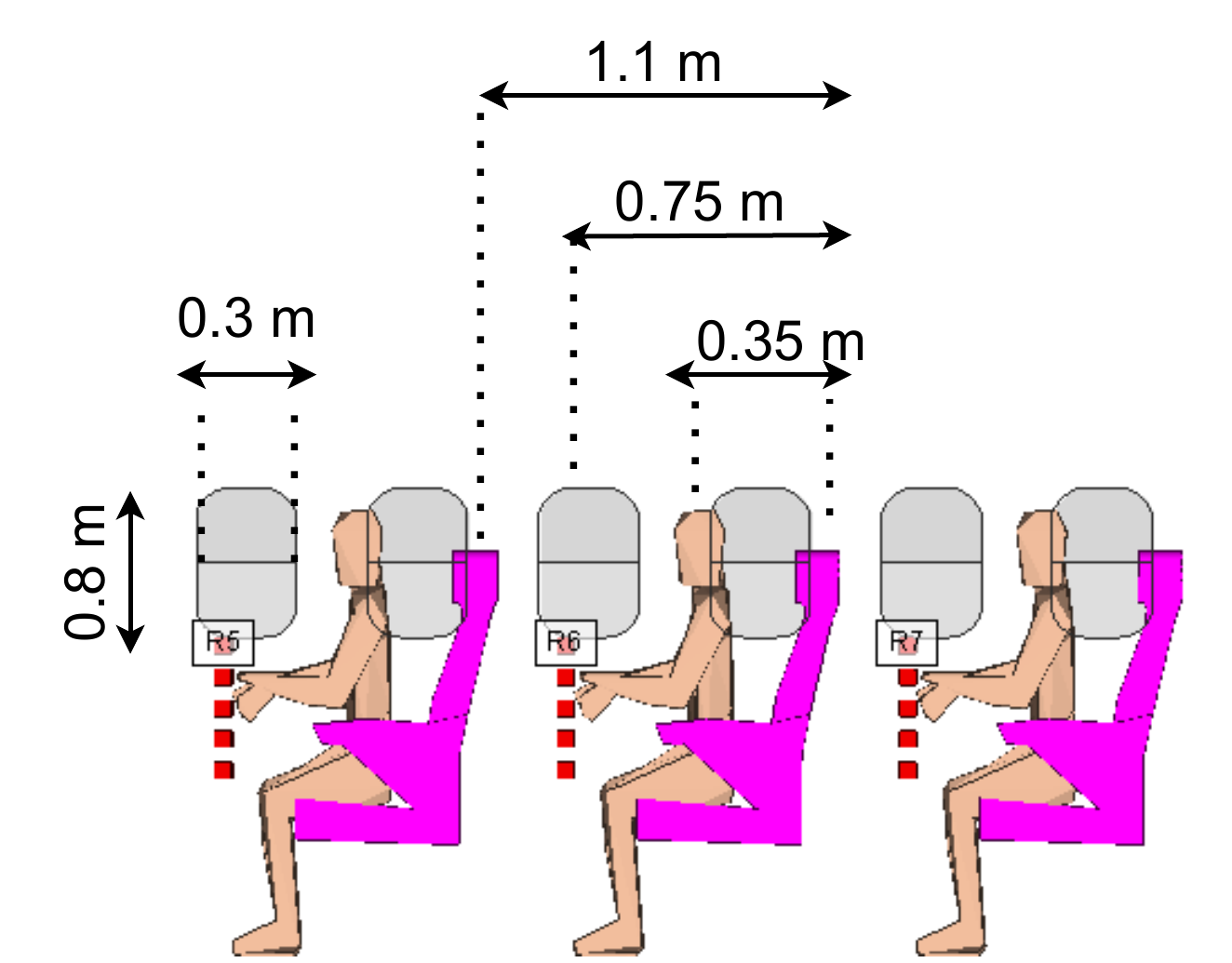} }
	\end{center}
	\vspace{-0.5cm}
	\caption{Illustration of the IDS RT environment and geometry.}
	\label{fig:ray_tracing}
\end{figure}

To have a fair comparison between the different scenarios, we consider the same transmitter and receiver locations, and equal total area of the simulation environment. We consider a single compartment of 12 rows with 6 seats each. The length, height, and width of the IDS and the objects within are given in Fig. \ref{fig:ray_tracing}.  Seats, humans, and object geometries are preserved in different simulation cases. 

As illustrated in Fig. \ref{fig:north}, the TX is located at front panel of the cabin, 2.1 m above the floor and 1.7 m away from the sidewalls of the cabin in all scenarios. The distance of the TX from the end side of the cabin is 13.5 m. As shown in Fig. \ref{fig:east}, RX points are distributed as vertical surfaces with 0.1 m seperation from one another. A vertical surface spans from 0.6 m to 1 m in the vertical dimension and spans all the width of the cabin. The dimensions are selected based on the minimum and the maximum heights of the possible hand movements of a human sitting on a seat. Each vertical RX surface level is 0.75 m away from the corresponding seat. We consider 12 vertical surfaces, 1 per row, which makes 2400 RX points in total. 

\subsection{EM Characteristics of Materials}
The material characteristics are critical in accurate modeling of the propagation environment. The dielectric parameters of the materials are given in Table \ref{tab:materials}. The passenger seats are modeled by nylons. Metal and carbon-glass composite are considered for fuselages in different simulation scenarios. For the windows, we consider glass as the closest material. In this table, $\epsilon$ denotes the complex permittivity. The majority of literature selects specific wagon or aircraft and calibrates the  materials in RT simulations according to that environment \cite{cabin_60,train_60}. In this work, the material characteristics are obtained from several different measurement works considering the given frequency bands in their analyses. 
\begin{table}[tb]
	\centering
\caption{Dielectric Properties of Materials Considered in the Simulations at 28 GHz.}
\begin{tabular}{|l|l|l|}
\hline
Material    & $\epsilon$ & Thickness (cm) \\ \hline
Metal (PEC) &  1      & n/a                    \\ \hline
Glass-Carbon composite \cite{gfrc} &  4.50-0.05j    &  0.3                          \\ \hline
Human skin \cite{human_tissue}  &  19.3-19.5j           & 0.1            \\ \hline
Nylon \cite{nylon}   &    3.01-0.021j         &    0.25                \\ \hline
Glass \cite{itu} &     6.27-0.1469j       & 0.3     \\ \hline
\end{tabular}
\label{tab:materials}
\end{table}

     \vspace{-0.1cm}

\section{Channel Characteristics and Discussion}

In this section, we present results of the RT based channel evaluation for IDS at 28 GHz considering the four scenarios defined in Section \ref{sec:sim_scenarios}. By comparing channel characteristics for different scenarios and the related channel models, we discuss the influencing factors of the environment on different channel characteristics. During the analyses, we categorize the results into two different main cases, line of sight (LOS) and non-line of sight (NLOS). In the NLOS conditions, we also consider diffuse scattering (DS) region.  If the TX and an RX have a direct path that does not experience any blockage, then the RX is in LOS region. If direct path is not available, the RX is in the NLOS region. In the DS region, the receivers only have paths resulted from diffuse scattering.  The main parameters that will be discussed below are  $\mathrm{PL}$, $\mathrm{SF}$, $\mathrm{DS}$, $\mathrm{KF}$, $\mathrm{ASD}$, $\mathrm{ASA}$, $\mathrm{ESD}$, $\mathrm{ESA}$. 
For the enlisted parameters, we provide their average value and standard deviation  $\mu_{{\mathrm{par}}}$, and $\sigma_{{\mathrm{par}}}$ respectively for the parameter  $ \mathrm{par} \in \{\mathrm{PL},  \mathrm{SF} , \mathrm{DS} , \mathrm{KF} , \mathrm{ASD} , \mathrm{ASA} , \mathrm{ESD} , \mathrm{ESA} \}$.

 \subsection{Propogation Cases and Received Signal Strengths}
 \begin{figure}[tb]
 \centering
\includegraphics[width=0.75\linewidth]{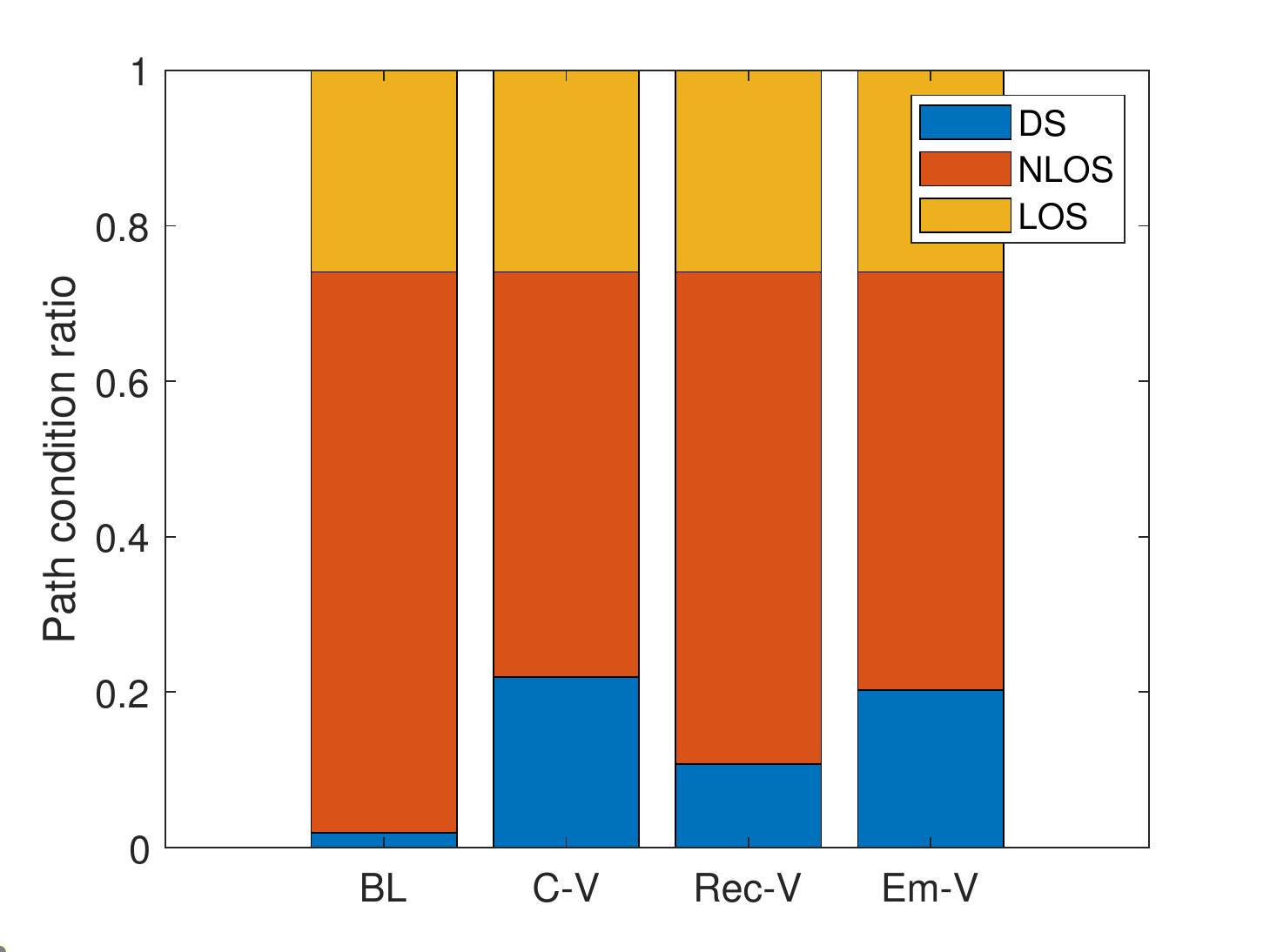}
     \vspace{-0.5cm}
\caption{The ratios of propagation cases in different simulation scenarios.}	\label{fig:regions}
\end{figure}
\begin{figure}[tb]
\includegraphics[width=0.9\linewidth]{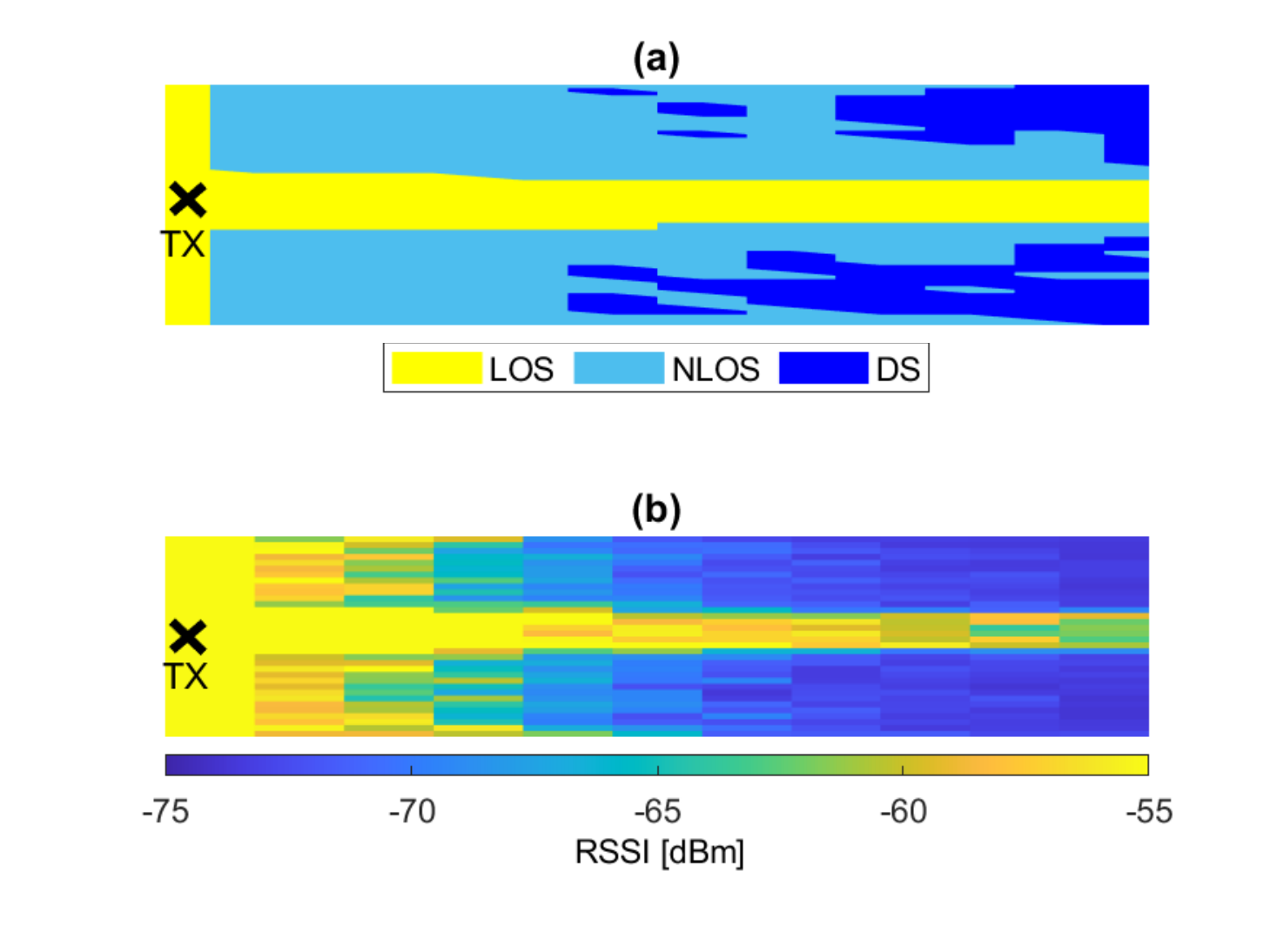}
     \vspace{-0.5cm}
\caption{(a) Received signal path characteristics and (b) RSSI levels for the C-V cases when the RX height is set to 0.7. DS denotes diffuse scattering.}	\label{fig:RSSI_vs_cases}
\end{figure}
Fig. \ref{fig:regions} shows the ratios of the path conditions for the considered scenarios. One observation is that the LOS ratio is similar in all scenarios. This can be explained by the Fig. \ref{fig:RSSI_vs_cases}(a), where the LOS condition is only accessible by the RX points in the front row  and the corridor of the fuselage. Another observation is that the DS condition has the lowest ratio in the  BL scenario. Since BL assumes metal fuselage, and densely populated by passengers, more RXs can get a path directly from reflections and diffractions instead of diffuse scattering. Reduced amount of NLOS in C-V and Em-V indicates the importance of the reflective material and the reflective advantage  of the human skin.

We present a 2D top view of an IDS C-V scenario in Fig. \ref{fig:RSSI_vs_cases}, with the TX located on the center left. Path conditions grouped in 3 are shown in Fig. \ref{fig:RSSI_vs_cases}(a) with the corresponding received signal strength indicator (RSSI) values in Fig. \ref{fig:RSSI_vs_cases}(b). We notice that the highest RSSI levels are observed in LOS conditions, where approximately 10 times higher RSSI levels are observed than in the NLOS case in the close locations. Also, due to the seats and passengers, we observe that under the NLOS condition, the signal severely attenuates as the TX-RX separation increases. However, when the TX-RX separation is short (for this example approximately $4$ m), we can see that NLOS users can still get acceptable signal strengths.  Under DS conditions, the RSSI values are critically reduced, showing the low contribution of the diffuse scattering to the received signal. 

\subsection{Path Loss and Shadowing}

\begin{figure*}[tb]
	\begin{center}	
		\subfigure[]{
			\label{fig:metal_PL}
			\includegraphics[width=0.255\textwidth]{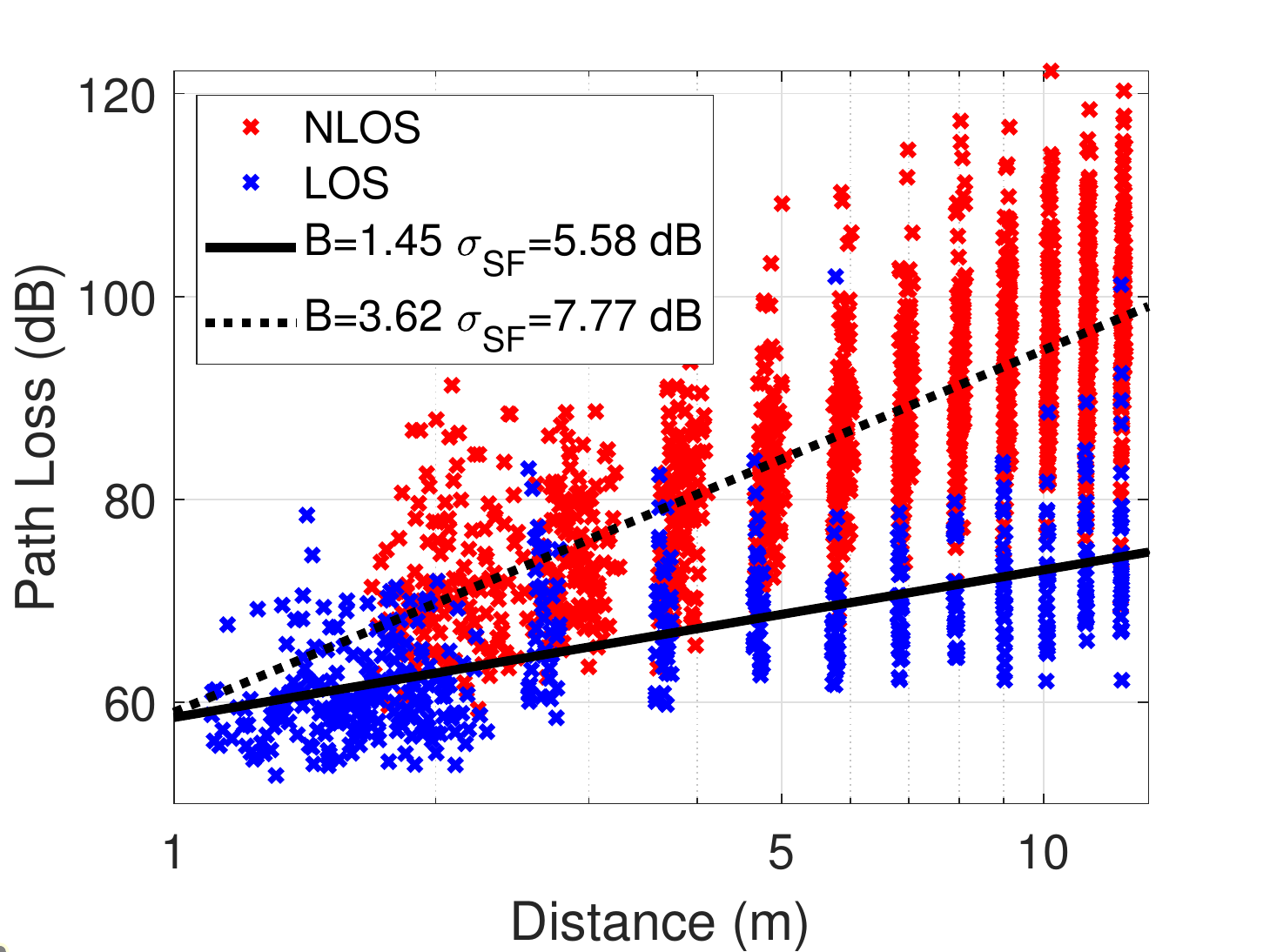} }
	\hspace{-0.7cm}
		\subfigure[]{
			\label{fig:carbon_PL}
			\includegraphics[width=0.255\textwidth]{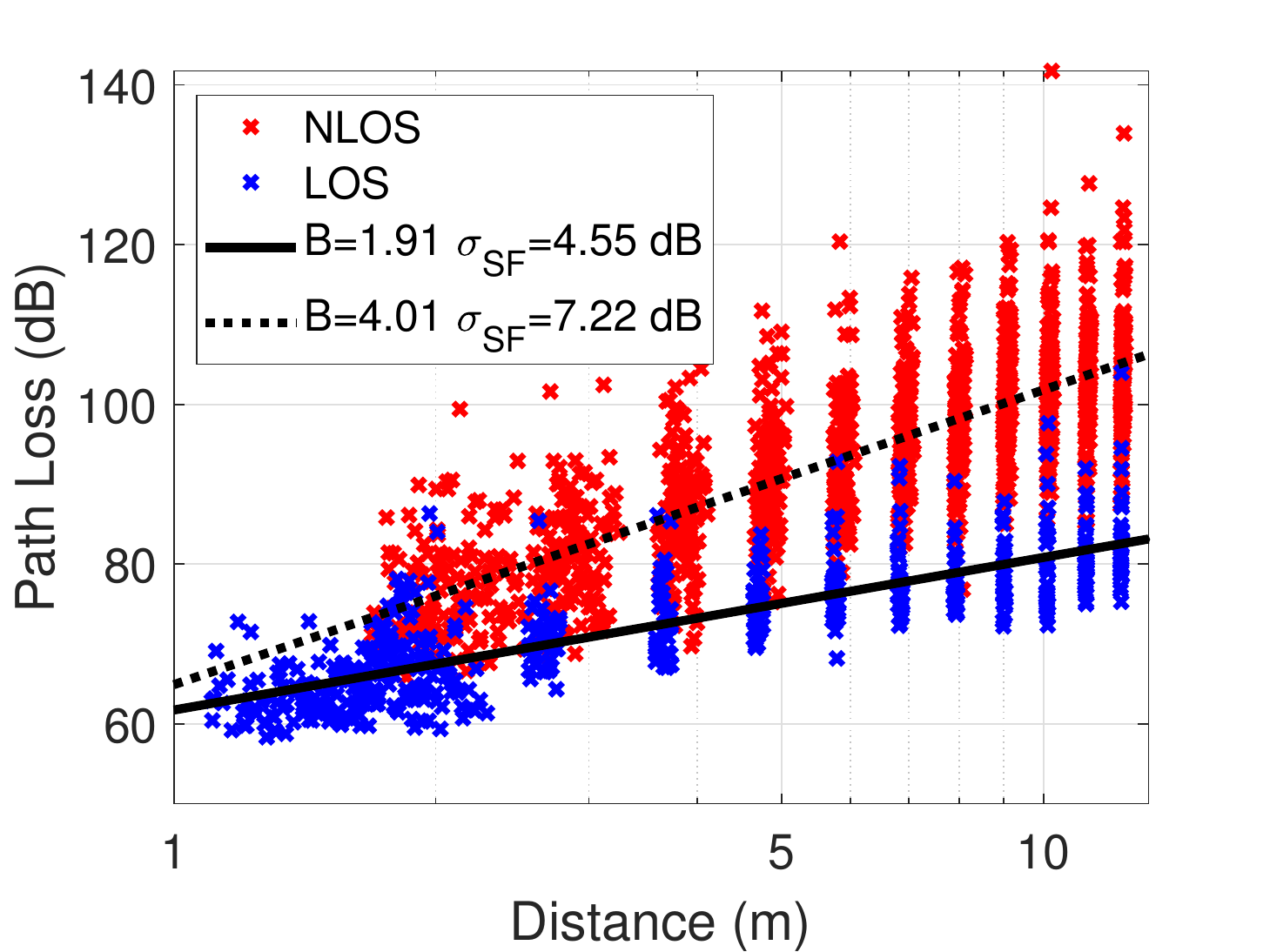} }
			\hspace{-0.7cm}
		\subfigure[]{
			\label{fig:rectangular_PL}
			\includegraphics[width=0.255\textwidth]{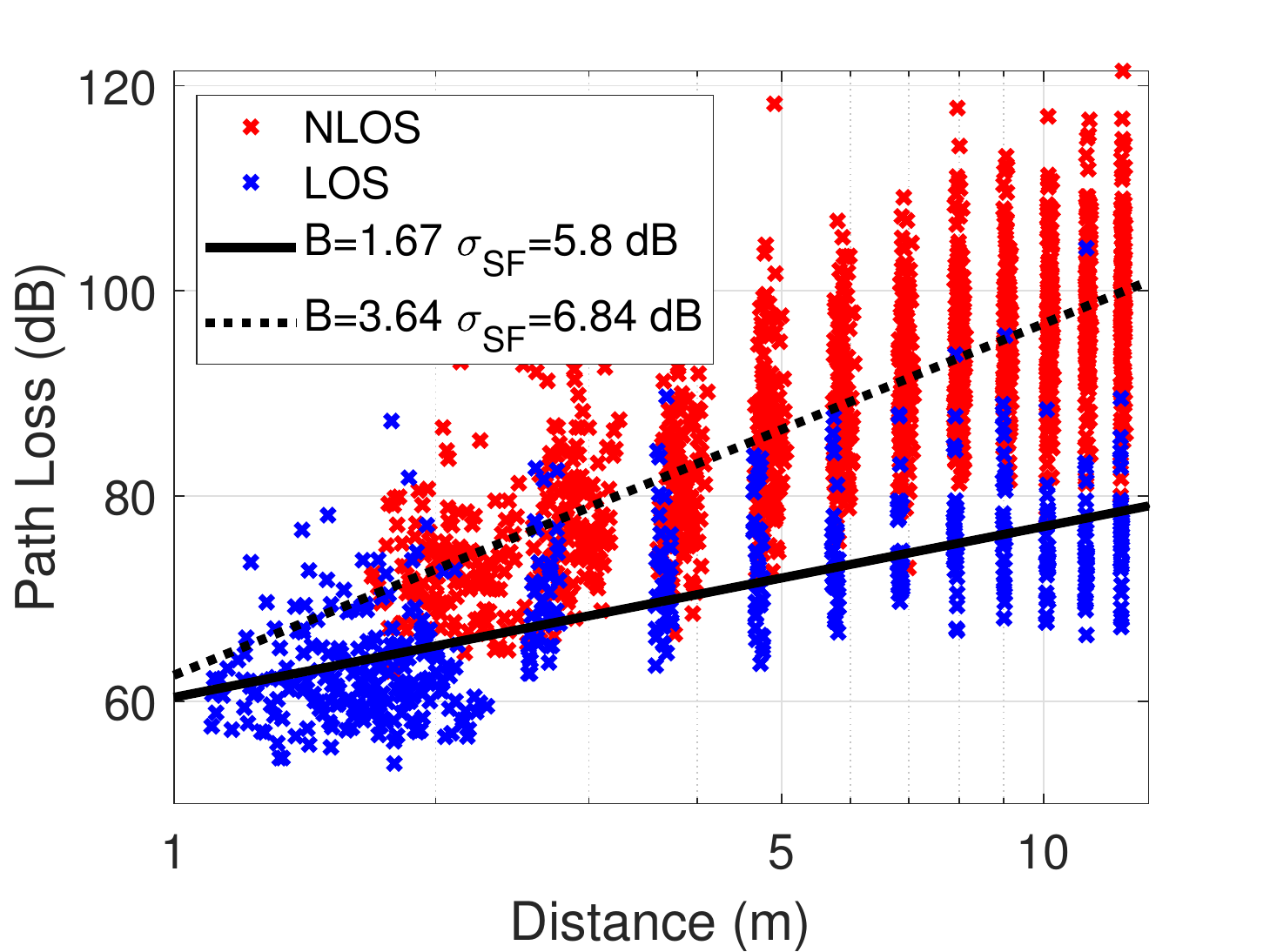} }
		\hspace{-0.7cm}
		\subfigure[]{
			\label{fig:empty_PL}
			\includegraphics[width=0.255\textwidth]{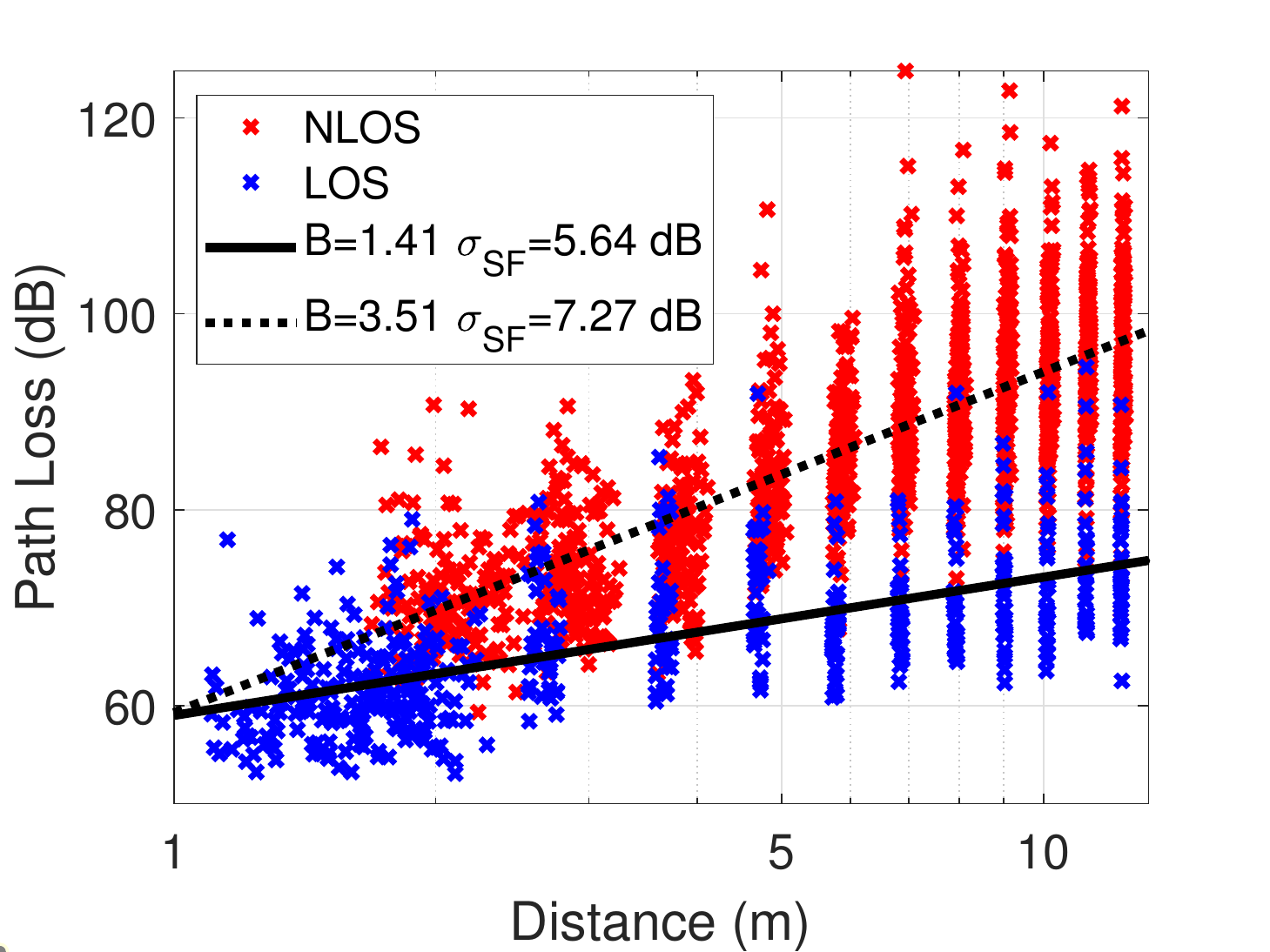} }
	\end{center}
	\vspace{-0.5cm}
	\caption{Path loss for RT simulation results and fitted curve considering (a) BL, (b) C-V, (c) Rec-V, (d) Em-V. }
	\label{fig:PL}
\end{figure*}
We will analyze the  path loss and shadowing, and  compare  with the existing 3GPP and the state-of-the-art measurement work \cite{3gpp, train_28}. As in these works, we consider an  A-B path loss model that can be represented by

\begin{equation}
\text{PL [dB]} = A +10 B\log_{10}\left(\frac{d}{d_0}\right)+ \chi_\sigma,
\end{equation}
where $\text{PL}$ denotes the path loss. ${A}$ is the experienced path loss at the reference distance $d_0$. $B$ is the linear slope (the average path loss exponent), and $d$ denotes the 3-dimensional (3-D) distance between TX and RX. $d_0=1$ m is considered as the reference distance.  $\chi_\sigma$ is the shadow fading, which is modeled as a normal random variable in dB scale with zero mean and $\sigma_{\text{SF}}$ standard deviation.

\textcolor{black}{In Fig. \ref{fig:PL}, lines illustrate the fitted path loss curves, while crosses show path loss results from RT simulations. The path loss curves are fitted by MATLAB's curve fitting tool. The dB difference of the RT simulation results from the fitted path loss curves are fitted to a normal random distribution by MATLAB's distribution fitting tool. $\sigma_{\mathrm{SF}}$ values are obtained as the standard deviation of the fitted distribution.} As expected, RXs in NLOS case, experience larger path losses than the RXs in the LOS case. \textcolor{black}{Similar to the other indoor channel modeling works, the $A$ parameters are around 59-62 dB,} and  the $B$ parameters for LOS cases are lower than 2. For the NLOS cases, the highest $B$ parameter is observed in C-V, which shows the disadvantageous effect of using non-reflective material. The remaining parameters are similar for each scenario, showing that the path loss and shadowing can be modeled with a single parameter set in various IDS environments.

A comparison with the relevant previous indoor channel modeling approaches is provided in Table \ref{tab:all_results_parameters}. 
The similarity in path loss characteristics between our work and  \cite{train_28} shows that our RT simulation results are in line with the measurement analysis. Another interesting comparison is with the 3GPP IO channel model.  In comparison to 3GPP IO, IDS contains a dense population of humans, and seats which have both advantages and disadvantages. As a disadvantage, the large number of blocking objects degrades the overall performance in the NLOS cases. As an advantage, the IDS fuselage and object materials are more reflective than the regular indoor walls and objects. The similarity of the $A$, $B$, and $\sigma_{SF}$ values between 3GPP IO and our model shows that the advantages and disadvantages in IDS environment balances each other out, and provides similar large-scale fading characteristics with indoor office spaces at the end.  

\subsection{Rician K-Factor}
\begin{table*}[tb]
\centering
\caption{Channel parameters of Indoor Dense Spaces with different material, denseness, geometry cases at 28 GHz.}
\begin{tabular}{|l|l|cc||cc||cc||cc||cc||cc|}
\hline
Param. Type & IDS Scenarios            & \multicolumn{2}{l||}{BL}                    & \multicolumn{2}{l||}{C-V }                   & \multicolumn{2}{l||}{Rec-V}            & \multicolumn{2}{l||}{Em-V}                   & \multicolumn{2}{l||}{3GPP InO}                          & \multicolumn{2}{l|}{IW \cite{train_28}} \\ \cline{2-14}
& Condition          & \multicolumn{1}{l|}{LOS}   & \multicolumn{1}{l||}{NLOS} & \multicolumn{1}{l|}{LOS}   & \multicolumn{1}{l||}{NLOS} & \multicolumn{1}{l|}{LOS} & \multicolumn{1}{l||}{NLOS} & \multicolumn{1}{l|}{LOS}   & \multicolumn{1}{l||}{NLOS} & \multicolumn{1}{l|}{LOS}   & \multicolumn{1}{l||}{NLOS} & \multicolumn{1}{l|}{LOS}         & \multicolumn{1}{l|}{NLOS$^1$}       \\ \hline
\multirow{3}{*}{Large-scale} &A  {[}dB{]}               & \multicolumn{1}{c|}{58.49} & 59.00                      & \multicolumn{1}{c|}{61.72} & 64.90                     & \multicolumn{1}{c|}{60.37}    &  62.52                         & \multicolumn{1}{c|}{59.00} & 59.28                     & \multicolumn{1}{c|}{61.34} & 53.33                     & \multicolumn{1}{c|}{59.04}       & 52.24                            \\ \cline{2-14}
&B {[}dB{]}                & \multicolumn{1}{c|}{1.45}  & 3.62                      & \multicolumn{1}{c|}{1.91}  & 4.00                      & \multicolumn{1}{c|}{1.66}    &  3.64                         & \multicolumn{1}{c|}{1.41}  & 3.94                      & \multicolumn{1}{c|}{1.73}  & 3.83                      & \multicolumn{1}{c|}{1.48}        & 3.50                             \\ \cline{2-14}
&$\sigma_{SF}$ {[}dB{]}    & \multicolumn{1}{c|}{5.58}   & 7.76                       & \multicolumn{1}{c|}{4.55}  & 7.22                     & \multicolumn{1}{c|}{5.80}    &  6.84                         & \multicolumn{1}{c|}{5.64}  & 9.38                      & \multicolumn{1}{c|}{3}     & 8.03                      & \multicolumn{1}{c|}{1.41}        & 7.26                            \\ \cline{1-14}
\multirow{4}{*}{Small-scale} & $\mu_{\mathrm{KF}}$ {[}dB{]}       & \multicolumn{1}{c|}{-4.51}  & n/a                        & \multicolumn{1}{c|}{4.3}  &   n/a                       & \multicolumn{1}{c|}{-1.98}    & n/a                       & \multicolumn{1}{c|}{-4.81}  & n/a                       & \multicolumn{1}{c|}{7}     & n/a                       & \multicolumn{1}{c|}{-}            & n/a                              \\ \cline{2-14}
&$\sigma_{\mathrm{KF}}$ {[}dB{]}    & \multicolumn{1}{c|}{-8.11}  & n/a                       & \multicolumn{1}{c|}{2.32}  & n/a                       & \multicolumn{1}{c|}{-4.69}    & n/a                       & \multicolumn{1}{c|}{-8.07}  & n/a                       & \multicolumn{1}{c|}{4}     & n/a                       & \multicolumn{1}{c|}{-}            & n/a                              \\ \cline{2-14}
&$\mu_{\mathrm{DS}}$ {[}ns{]}       & \multicolumn{1}{c|}{5.60}  & 3.82                      & \multicolumn{1}{c|}{2.58}  & 2.60                      & \multicolumn{1}{c|}{11.82}      &  5.68                         & \multicolumn{1}{c|}{5.92}  & 4.70                      & \multicolumn{1}{c|}{19.65}   &   26.15                      & \multicolumn{1}{c|}{-}            &    -                              \\ \cline{2-14}
&$\sigma_{\mathrm{DS}}$ {[}ns{]}    & \multicolumn{1}{c|}{2.35}  & 2.45                     & \multicolumn{1}{c|}{1.46}  & 1.83                      & \multicolumn{1}{c|}{5.56}      &  3.22                          & \multicolumn{1}{c|}{2.50}  & 3.22                      & \multicolumn{1}{c|}{ 1.51e8}   &  1.58e8                      & \multicolumn{1}{c|}{-}            &    -                              \\ \hline
\multirow{8}{*}{Angular} &$\mu_{\mathrm{ASD}}$ {[}deg.{]}    & \multicolumn{1}{c|}{39.02} & 15.88                     & \multicolumn{1}{c|}{18.12} & 6.32                      &  \multicolumn{1}{c|}{11.87}      & 3.80                      & \multicolumn{1}{c|}{38.67} & 15.30                     & \multicolumn{1}{c|}{39.81}   &  41.68                      & \multicolumn{1}{c|}{-}            &  -                                \\ \cline{2-14}
&$\sigma_{\mathrm{ASD}}$ {[}deg.{]} & \multicolumn{1}{c|}{15.35} & 10.51                     & \multicolumn{1}{c|}{12.63} & 6.19                      &\multicolumn{1}{c|}{7.02}      &  3.61                        & \multicolumn{1}{c|}{ 15.12}  & 10.96                     & \multicolumn{1}{c|}{1.51}   &  1.72                      & \multicolumn{1}{c|}{-}            &   -                               \\ \cline{2-14}
&$\mu_{\mathrm{ASA}}$ {[}deg.{]}      & \multicolumn{1}{c|}{39.25} & 19.18                     & \multicolumn{1}{c|}{15.36} & 14.25                      & \multicolumn{1}{c|}{29.80}      & 14.35                      & \multicolumn{1}{c|}{41.66} & 13.61                     & \multicolumn{1}{c|}{31.85}   & 50.36                         & \multicolumn{1}{c|}{-}         &              -                    \\ \cline{2-14}
&$\sigma_{\mathrm{ASA}}$ {[}deg.{]}   & \multicolumn{1}{c|}{15.10} & 11.02                     & \multicolumn{1}{c|}{10.11}  & 7.49                      &\multicolumn{1}{c|}{17.19}      & 8.23                           & \multicolumn{1}{c|}{9.81}  & 13.60                     & \multicolumn{1}{c|}{1.97}   & 1.71                         & \multicolumn{1}{c|}{-}           &   -                               \\ \cline{2-14}
&$\mu_{\mathrm{ESD}}$ {[}deg.{]}      & \multicolumn{1}{c|}{31.66} & 16.34                     & \multicolumn{1}{c|}{22.87} & 11.33                      & \multicolumn{1}{c|}{70.24}      &  30.31                       & \multicolumn{1}{c|}{31.48} & 17.51                     & \multicolumn{1}{c|}{1.37}   &  12.02                       & \multicolumn{1}{c|}{-}          &           -                       \\ \cline{2-14}
&$\sigma_{\mathrm{ESD}}$ {[}deg.{]}   & \multicolumn{1}{c|}{39.97} & 12.39                     & \multicolumn{1}{c|}{25.33} & 6.76                      & \multicolumn{1}{c|}{30.21}      & 18.03                        & \multicolumn{1}{c|}{39.23} & 14.18                     & \multicolumn{1}{c|}{3.09}   &   2.29                       & \multicolumn{1}{c|}{-}           &  -                                \\ \cline{2-14}
&$\mu_{\mathrm{ESA}}$ {[}deg.{]}      & \multicolumn{1}{c|}{50.18} & 72.83                     & \multicolumn{1}{c|}{35.36} & 70.33                     & \multicolumn{1}{c|}{61.72}      &  70.68                         & \multicolumn{1}{c|}{52.01} & 70.18                     & \multicolumn{1}{c|}{11.47}   &  14.71                     & \multicolumn{1}{c|}{-}            &   -                               \\ \cline{2-14}
&$\sigma_{\mathrm{ESA}}$ {[}deg.{]}   & \multicolumn{1}{c|}{25.40} & 26.63                     & \multicolumn{1}{c|}{18.65} & 28.51                     &  \multicolumn{1}{c|}{19.44}      & 28.15                        & \multicolumn{1}{c|}{25.55} & 28.36                    & \multicolumn{1}{c|}{1.60}   &  4.11                         & \multicolumn{1}{c|}{-}        &    -                             \\ \hline

\multicolumn{12}{ l }{$^1$This case only considers limited hand blockage.}  \\
\hline

\end{tabular}

\label{tab:all_results_parameters}
\end{table*}
$\mathrm{KF}$ shows the  ratio of the power coming from the strongest path and the sum of the power coming from all other paths. It indicates how strong the scattering is in the environment. In this work, we calculate $ \mathrm{KF}$ by dividing the power incoming from the direct path to the other available paths. Table \ref{tab:all_results_parameters} gives the mean values, $\mu_{\mathrm{KF}}$ and the standard deviations, $\sigma_{\mathrm{KF}}$ of the $\mathrm{KF}$s obtained for the considered scenarios.

Within all of the results, the $ \mathrm{KF}$ is the most varying parameter in different scenarios and in comparisons to the available cases. We observe the lowest $ \mathrm{KF}$ values, around $-4.5$ dB, in the metal fuselage BL and Em-V scenarios. This effect shows the importance of the material in the multipath richness of the environment. Since metal is the most reflective material available, we can observe higher number of strong reflections in the environment coming from different paths. On the other hand, for C-V, we observe the highest $ \mathrm{KF}$, 4.3 dB, which indicates the multipath degradation resulted from absorbing materials. Another important observation is that all of the $ \mathrm{KF}$ values in our work are smaller than the 3GPP indoor office channel model. This illustrates the multipath richness resulted from the dense human and objects in IDS.

\subsection{Delay Spread}
In wideband channel modeling efforts, the delay spread (DS) is an important factor illustrating the multipath characteristics of the wireless channel. The main interest of the DS is the root mean square (RMS) value for having a sense of time dispersion in the wireless channel. RMS-DS is calculated as the second-order central
moment of the power delay profile by
\begin{equation}
    \sigma_{\tau}=\sqrt{\frac{\sum_{n=1}^{N} \tau_{n}^{2}  P_{n}}{\sum_{n=1}^{N} P_{n}}-\left(\frac{\sum_{n=1}^{N} \tau_{n}  P_{n}}{\sum_{n=1}^{N} P_{n}}\right)^{2}},
\end{equation}
where $\sigma_{\tau}$ denotes the RMS-DS \cite{Rappaport}. $P_n$ and $\tau_n$ denote respectively the received power and the excess time from the $n^{\mathrm{th}}$ multipath component. The total number of multipath components for a receiver is denoted by $N$, where this is a variable that can be changed at each receiving location. Since we consider different receiving locations in the simulations, similar to the $ \mathrm{KF}$ analysis, we provide the mean, $\mu_{\mathrm{DS}}$ and the standard deviation, $\sigma_{\mathrm{DS}}$, for the RMS-DS in  Table \ref{tab:all_results_parameters}. For all considered scenarios, the mean RMS-DS values are around $2$-$6$ ns., which is almost $60\%$ lower than in the 3GPP indoor office model \cite{3gpp}. This is a direct result of the short distances in IDS in comparison to the indoor office models. Even though the incoming paths experience reflections and diffractions, the shorter distances forces them to be received at almost the same time. One interesting result is that the LOS condition has higher RMS-DS values than the NLOS condition in the BL, Rec-V, and Em-V scenarios. The LOS cases have fast direct paths and a collection of slower scattered paths, while the NLOS conditions only have the collection of slower scattered paths. Therefore, the delay spread is reduced due to the unavailability of the direct paths in NLOS cases.

\subsection{Angular Spread}
Angular spread analysis is another important indicator of the multipath richness of the wireless channel, and gives an overview on the analog beamforming capability. \textcolor{black}{For example, if the signal is received from a narrower angle, then TX can direct the signal towards that angle to concentrate the power at the RX.}  Four different angular properties of the IDS channel are analyzed with the 3GPP methodology from \cite{3gpp} that is described below. First, the mean angle of the related characteristics is calculated as 
\begin{equation}
\nu_{\Theta}=\frac{\sum_{n=1}^{N} \theta_{n}  P_{n}}{\sum_{n=1}^{N} P_{n}}, 
\end{equation}
where $\nu_{\theta}$ is the mean angle value of the multipaths for a given receiver location. $\Theta$ can be the $\mathrm{ASD}$, $\mathrm{ASA}$, $\mathrm{ESD}$, $\mathrm{ESA}$ according to the relevant analysis. $\theta_n$ denotes the angle of the $n^{\text{th}}$ multipath, and $\theta_n\in (-\pi,\pi]$. The second step is calculating the difference between the incoming angle and the mean angle in radian as
\begin{equation}
\theta_{n, \mu}=\bmod \left(\theta_{n}-\nu_{\theta}+\pi, 2 \pi\right)-\pi.
\end{equation}
As a final step, the RMS angle value is calculated as
\begin{equation}
\sigma_{AS}=\sqrt{\frac{\sum_{n=1}^{N}\left(\theta_{n, \mu}\right)^{2}  P_{n}}{\sum_{n=1}^{N} P_{n}}}.
\end{equation}
In Table \ref{tab:all_results_parameters}, we provide the mean and standard deviation values of the different receiving locations for the RMS angular spread values. In all scenarios, we observe lower $\mathrm{ASA}$ and $\mathrm{ASD}$ values in NLOS cases. This indicates that reflected and diffracted signals come from fewer different directions in NLOS cases compared to LOS cases. Another important observation is that the angular spread values are higher in BL, Rec-V and Em-V compared to C-V. This implies that highly reflective fuselage would create a large number of reflections arriving from a diverse set of angles. The channel parameters presented in Table III can be utilized in commercial channel generation tools such as QuaDRiGa \cite{quadriga} to realize IDS at 28 GHz.

\subsection{Bit Error Rate Comparison }
\begin{figure}[tb]
   \centering
   \includegraphics[width=0.85\linewidth]{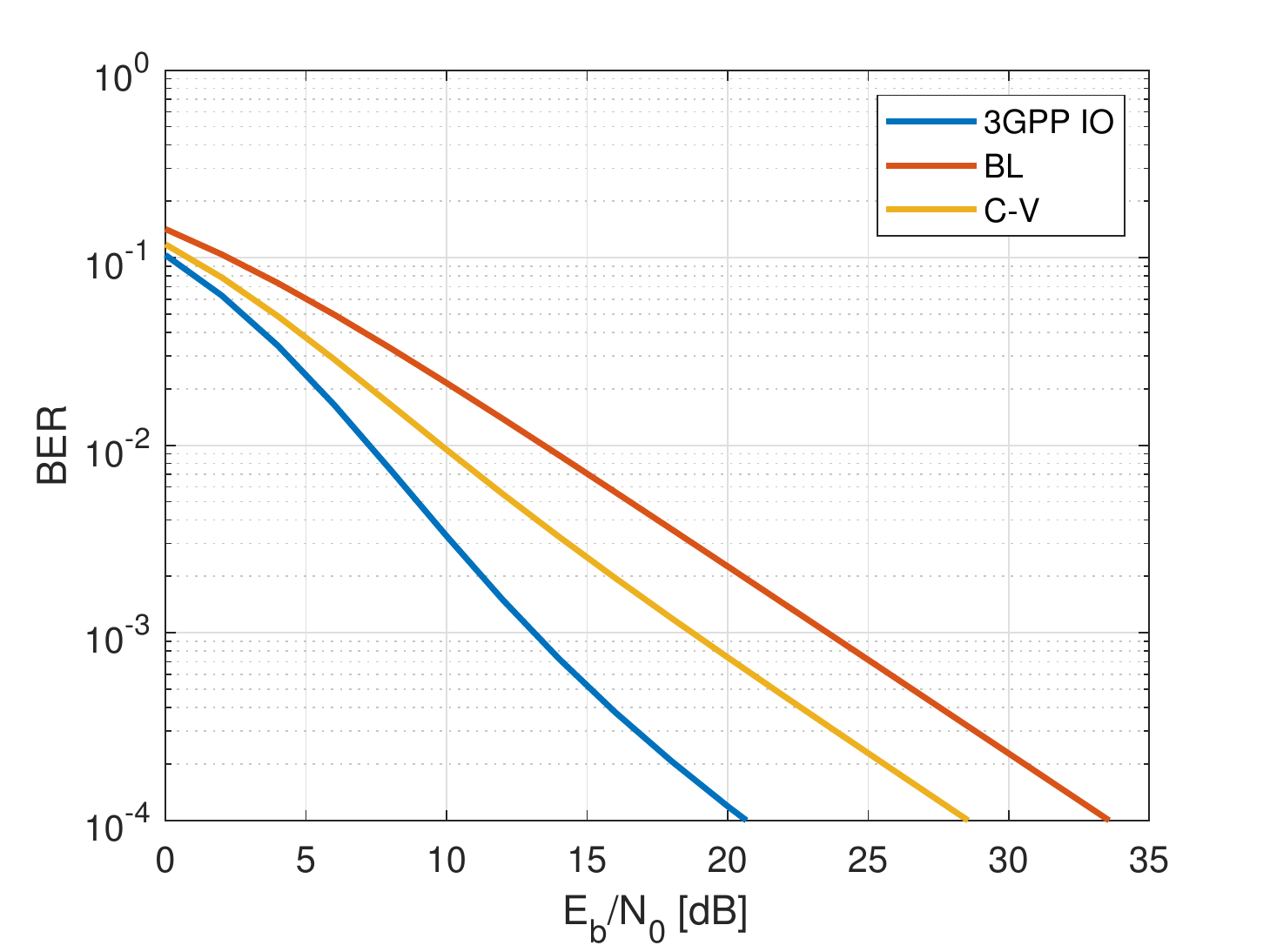}
         \vspace{-0.4cm}
   \caption{BER comparison for different channel cases considering BPSK modulation.}
   \label{fig:BER}
     \vspace{-0.5cm}

\end{figure}

In this part, we compare the BER performance of the proposed channel models for IDS in the different scenarios with the 3GPP indoor channel model. We consider uncoded binary-phase-shift keying (BPSK) modulation for the sake of the simplicity. As it can be seen from Fig. \ref{fig:BER}, the 3GPP model constitutes an optimistic BER performance for the same signal energy level. Since the $ \mathrm{KF}$ is lower in IDS scenarios compared to the 3GPP model, small-scale fading results in  worse BER performance. For example, for the targeted $10^{-3}$ BER, the receiver requires 10 times higher signal power in the BL scenario compared to the 3GPP model. Therefore, 3GPP IO channel model is not applicable to IDS environment, and our channel characterization is more effective in illustrating the rich multipath scattering effect in IDS.

\section{Conclusion}
The millimeter wave (mmWave) signal propagation at 28 GHz is analyzed for confined indoor environments containing dense users and objects, such as aircrafts, train wagons, and hyperloops. Terming this environment as indoor dense spaces (IDS),  we observe  strong received signal strength levels for  non-line-of-sight (NLOS) cases (5 dB lower than line-of-sight) for shorter than 4-5 meters TX-RX seperation, indicating that NLOS access can be provided in short-range mmWave systems. While path loss and shadowing  characteristics are similar to the 3GPP indoor office (IO) channel model, at least 3 dB smaller Rician-K parameter and $60\%$ lower mean delay spread values in IDS imply much more scattered and reflected paths arriving at similar times due to the confined geometry, and dense objects. The electromagnetic (EM) characteristics  of the cabin material make the strongest impact on small-scale channel parameters compared to  the geometry and human presence.
     \vspace{-0.3cm}

\section*{Acknowledgment}
Results incorporated in this paper received funding from the ECSEL Joint Undertaking (JU) under grant
agreement No 876124. The JU receives support from the EU Horizon 2020 research and
innovation programme and Vinnova in Sweden.

\balance
\bibliographystyle{IEEEtran}

\bibliography{references}

\begin{thebibliography}{10}
\providecommand{\url}[1]{#1}
\csname url@samestyle\endcsname
\providecommand{\newblock}{\relax}
\providecommand{\bibinfo}[2]{#2}
\providecommand{\BIBentrySTDinterwordspacing}{\spaceskip=0pt\relax}
\providecommand{\BIBentryALTinterwordstretchfactor}{4}
\providecommand{\BIBentryALTinterwordspacing}{\spaceskip=\fontdimen2\font plus
\BIBentryALTinterwordstretchfactor\fontdimen3\font minus
  \fontdimen4\font\relax}
\providecommand{\BIBforeignlanguage}[2]{{%
\expandafter\ifx\csname l@#1\endcsname\relax
\typeout{** WARNING: IEEEtran.bst: No hyphenation pattern has been}%
\typeout{** loaded for the language `#1'. Using the pattern for}%
\typeout{** the default language instead.}%
\else
\language=\csname l@#1\endcsname
\fi
#2}}
\providecommand{\BIBdecl}{\relax}
\BIBdecl

\bibitem{ericsson}
Ericsson, ``{Ericsson Mobility Report, Mobile subscriptions Q2 2021},'' August
  2021.

\bibitem{Rappaport_mmWave}
G.~R. Maccartney, T.~S. Rappaport, S.~Sun, and S.~Deng, ``Indoor office
  wideband {Millimeter-Wave} propagation measurements and channel models at 28
  and 73 {GHz} for ultra-dense {5G} wireless networks,'' \emph{IEEE Access},
  vol.~3, pp. 2388--2424, 2015.

\bibitem{factory}
D.~Solomitckii, A.~Orsino, S.~Andreev, Y.~Koucheryavy, and M.~Valkama,
  ``Characterization of {mmWave} channel properties at 28 and 60 {GHz} in
  factory automation deployments,'' in \emph{IEEE Wireless Communications and
  Networking Conf.}, 2018, pp. 1--6.

\bibitem{cabin_60}
R.~Felbecker, W.~Keusgen, and M.~Peter, ``Incabin millimeter wave propagation
  simulation in a wide-bodied aircraft using {Ray-Tracing},'' in \emph{IEEE
  68th Vehicular Technology Conference}, 2008, pp. 1--5.

\bibitem{train_28}
L.~Rubio, V.~M. Rodrigo~Peñarrocha, J.-M. Molina-García-Pardo,
  L.~Juan-Llácer, J.~Pascual-García, J.~Reig, and C.~Sanchis-Borras,
  ``Millimeter wave channel measurements in an intra-wagon environment,''
  \emph{IEEE Trans. Veh. Technol.}, vol.~68, no.~12, pp. 12\,427--12\,431,
  2019.

\bibitem{ieee_standard}
A.~Maltsev, A.~Pudeyev, A.~Lomayev, and I.~Bolotin, ``Channel modeling in the
  next generation mmwave {Wi-Fi: IEEE 802.11ay} standard,'' in \emph{European
  Wireless Conference}, 2016, pp. 1--8.

\bibitem{3gpp}
``{5G, Study on Channel Model for Frequencies From 0.5 to 100 GHz (3GPP TR
  38.901 Version 16.1.0 Release 16)},'' Tech. Rep., 2020.

\bibitem{train_60}
K.~Guan, B.~Peng, D.~He, J.~M. Eckhardt, S.~Rey, B.~Ai, Z.~Zhong, and
  T.~Kürner, ``Channel characterization for intra-wagon communication at 60
  and 300 {GHz} bands,'' \emph{IEEE Trans. Veh. Technol.}, vol.~68, no.~6, pp.
  5193--5207, 2019.

\bibitem{cabin_measure_60}
M.~Peter, W.~Keusgen, A.~Kortke, and M.~Schirrmacher, ``Measurement and
  analysis of the 60 {GHz} in-vehicular broadband radio channel,'' in
  \emph{IEEE 66th Vehicular Technology Conference}, 2007, pp. 834--838.

\bibitem{bachmann2017environmental}
J.~Bachmann, C.~Hidalgo, and S.~Bricout, ``Environmental analysis of innovative
  sustainable composites with potential use in aviation {sector—A} life cycle
  assessment review,'' \emph{Science China Technological Sciences}, vol.~60,
  no.~9, pp. 1301--1317, 2017.

\bibitem{metro_28}
Y.~Xu, D.~He, H.~Yi, K.~Guan, M.~Heino, and M.~Sonkki, ``The influence of
  self-user shadowing in the intra-metro communication scenario at 28 {GHz},''
  in \emph{European Conference on Antennas and Propagation}, 2020, pp. 1--5.

\bibitem{gfrc}
A.~Tamburrano, F.~Marra, J.~Lecini, and M.~S. Sarto, ``Complex permittivity
  extraction method of a thin coating: {EM} properties of a {G}raphene-based
  film on a composite layer,'' in \emph{International Symposium on
  Electromagnetic Compatibility}, 2018, pp. 602--607.

\bibitem{human_tissue}
T.~Wu, T.~S. Rappaport, and C.~M. Collins, ``The human body and millimeter-wave
  wireless communication systems: {I}nteractions and implications,'' in
  \emph{International Conf. on Commun.}, 2015, pp. 2423--2429.

\bibitem{nylon}
B.~Riddle, J.~Baker-Jarvis, and J.~Krupka, ``Complex permittivity measurements
  of common plastics over variable temperatures,'' \emph{IEEE Trans. Microw.
  Theory Techn.}, vol.~51, no.~3, pp. 727--733, 2003.

\bibitem{itu}
I.~T.~U. Recs., ``Propagation data and prediction methods for the planning of
  indoor radio communication systems and radio local area networks in the
  frequency range {900 MHz to 100 GHz},'' 2012, {Geneva, Switzerland}.

\bibitem{Rappaport}
T.~S. Rappaport, \emph{{Wireless Communications: Principles and
  Practice}}.\hskip 1em plus 0.5em minus 0.4em\relax Englewood Cliffs, NJ:
  Prentice-Hall, 1996, vol.~2.

\bibitem{quadriga}
S.~Jaeckel, L.~Raschkowski, K.~Börner, and L.~Thiele, ``{QuaDRiGa}: A {3-D}
  multi-cell channel model with time evolution for enabling virtual field
  trials,'' \emph{IEEE Trans. Antennas Propag.}, vol.~62, no.~6, pp.
  3242--3256, 2014.

\end{thebibliography}

\end{document}